\documentclass[letterpaper, 10 pt, conference]{ieeeconf}  

\IEEEoverridecommandlockouts                              
\usepackage[utf8]{inputenc}
\usepackage[T1]{fontenc}
\usepackage{graphicx}
\usepackage{footmisc}
\usepackage{subfigure}
\usepackage{url}
\usepackage{multirow}
\usepackage[noadjust]{cite}
\usepackage{amsmath,amsthm}
\usepackage{amssymb,amsfonts}
\usepackage{booktabs}
\usepackage{color}
\usepackage{ccaption}
\usepackage{booktabs}
\usepackage{float}
\usepackage{fancyhdr}
\usepackage{caption}
\usepackage{xcolor,stfloats}
\usepackage{comment}
\graphicspath{{figures/}}
\usepackage{cuted}  
\usepackage{epstopdf}
\usepackage{algorithmic}
\usepackage{algorithm}
\usepackage{makecell}
\usepackage{amsmath,bm}
\usepackage{wrapfig}

\makeatletter
\newenvironment{breakablealgorithm}
  {
   \begin{center}
     \refstepcounter{algorithm}
     \hrule height.8pt depth0pt \kern2pt
     \renewcommand{\caption}[2][\relax]{
       {\raggedright\textbf{\ALG@name~\thealgorithm} ##2\par}%
       \ifx\relax##1\relax 
         \addcontentsline{loa}{algorithm}{\protect\numberline{\thealgorithm}##2}%
       \else 
         \addcontentsline{loa}{algorithm}{\protect\numberline{\thealgorithm}##1}%
       \fi
       \kern2pt\hrule\kern2pt
     }
  }{
     \kern2pt\hrule\relax
   \end{center}
  }
\makeatother

\title{\LARGE \bf
Towards Calibrating Financial Market Simulators\\ with High-frequency Data
}

\author{Peng Yang, Junji Ren, Feng Wang, Ke Tang$^*$
\thanks{Peng Yang, Junji Ren, and Ke Tang are with the Guangdong Provincial Key Laboratory of Brain-inspired Intelligent Computation, Department of Computer Science and Engineering, Southern University of Science and Technology, Shenzhen 518055, China. E-mails: yangp@sustech.edu.cn; 12432695@mail.sustech.edu.cn; tangk3@sustech.edu.cn.}
\thanks{Feng Wang is with the School of Computer Science, Wuhan University, Wuhan 430072, China. E-mail: fengwang@whu.edu.cn.}%
}

\begin{document}

\maketitle
\thispagestyle{empty}
\pagestyle{empty}

\begin{abstract}

The fidelity of financial market simulation is restricted by the so-called "non-identifiability" difficulty when calibrating high-frequency data. This paper first analyzes the inherent loss of data information in this difficulty, and proposes to use the Kolmogorov-Smirnov test (K-S) as the objective function for high-frequency calibration. Empirical studies verify that K-S has better identifiability of calibrating high-frequency data, while also leads to a much harder multi-modal landscape in the calibration space. To this end, we propose the adaptive stochastic ranking based negatively correlated search algorithm for improving the balance between exploration and exploitation. Experimental results on both simulated data and real market data demonstrate that the proposed method can obtain up to 36.0\% improvement in high-frequency data calibration problems over the compared methods.

\end{abstract}

\section{Introduction}
\label{s:introduction}
\noindent Simulation has been widely acknowledged as an effective means of exploring the endogenous mechanisms of complex systems\cite{36,37,38,39}. For financial markets, a typical complex system, many risk events (such as "flash crashes") have not been clearly understood, which continuously threatens the security of the financial system\cite{1}. To reveal their causes, the Santa Fe Institute proposed the idea of artificial stock market model, viewing the stock market through the perspective of complex systems\cite{34}. Nowadays, the financial market simulator (FMS) has been frequently studied as an important analyzing tool\cite{2,40}.

So far, the development of FMSs has been mainly based on the agent-based modeling (ABM)\cite{3,4}. ABM is a multi-agent based simulation model, which usually consists of a simulated exchange agent and multiple types of parameterized simulated trader agents. During the simulation, trader agents continuously submit orders to the exchange agent. The exchange agent continuously matches these orders to generate simulated data according to the real market trading rules. That assures the simulated data conforms to the statistical properties of the real financial market data.

Although various prior knowledge can be introduced to build FMSs, there is still uncertainty in the real market that cannot be accurately modeled. ABM-based simulators deal with uncertainty in the form of parameterized agents, where different parameter settings lead to different generation process of the simulated data. Hence, the simulator basically represents the general generation process of the financial market data. One usually wants to study a specific risk event that happened in history, and the simulator should be able to reproduce the time series of the specific event accurately enough. How can we efficiently determine the relevant parameters of the FMS for such specific event? This problem is called the calibration of simulators \cite{5}.

The calibration problem can be described in the follow way: \textit{Suppose it is observed that the real system generates the state time series data with a length of T time steps $\{\widehat{s}_t\}_{t=1}^T, \widehat{s}_t\in \mathbb{R}^n$. Suppose that the domain experts have built a simulator $M(w,T^{\prime}):\mathbb{R}^d\rightarrow \mathbb{R}^{n\times T^{\prime}}, T^{\prime}\in\mathbb{N^+}$. The data generation process is determined by parameters $w$ \cite{6,7,8}, where $w\in \mathbb{R}^d$. Suppose that there is an optimal parameter $w^*$ such that $M(w^*,T)=\{\widehat{s}_t\}_{t=1}^T$. What is the value of $w^*$?} Calibration of a simulator is the inverse problem of building the simulator, which aims to find an inverse function $M^{-1}:\mathbb{R}^{n\times T}\rightarrow\mathbb{R}^d$ such that $w^*=M^{-1}(\{\widehat{s}_t\}_{t=1}^T)$. 
Here, $M^{-1}$ is the calibrator. Note that, since $M$ is usually a computer program, it is hard to derive the calibrator analytically. Therefore, the literature generally obtains it approximately by solving the following optimization problem \cite{5,9,10}:

\begin{equation}
\normalsize
w^*=\mathop{\arg\min}\limits_{w}D(\{\widehat{s}_t\}_{t=1}^T,M(w,T)=\{s_t\}_{t=1}^T), \\
\label{eq1}
\end{equation}

\noindent where $D$ is a discrepancy metric, $M(w,T)=\{s_t\}_{t=1}^T$ is the time series data generated by simulator $M$ under certain parameter $w$. Representative calibration methods include methods of simulated moments (MSM) \cite{11,12} and likelihood estimation \cite{13,14}.

In general, these calibration methods have improved the simulator's ability of simulating specific historical data to a certain extent. However, several experiments have found that this type of work can only calibrate low-frequency market data \cite{15,16}. On the other hand, many major risk events may be related to the popular high-frequency trading \cite{17}. If high-frequency data cannot be effectively simulated, due to the small sample characteristics of market risk events, the causes behind them can be hard to analyze \cite{18}.

In the literature, what makes it difficult for previous works to calibrate high-frequency data is called non-identifiability \cite{19}. Specifically, Some candidate parameters in the parameter space of the above objective function are non-identifiable as they all have the same objective function value. Such phenomenon suggests that the simulated data synthesized based on these different parameters have the same distance to the given historical data. However, the actual situation is usually that these simulated data enjoys a high similarity when compared with low-frequency (such as daily frequency) sampling of the given historical data, but suffers from a large difference when compared in high-frequency (such as second frequency). In other words, the objective function cannot evaluate which candidate parameter can better simulate the given high-frequency historical data.

The reason is that these objective functions all need to estimate the summary statistics (such as moments) of the parameterized distributions of both data first \cite{20}, and then calculate the distances of these summary statistics. Since the summary statistics can be viewed as a sort of dimension reduction of the original data, the detailed information of the original data will be lost. For the data to be calibrated, the higher its essential frequency is, the greater its detailed information will lose while calculating the summary statistics, making the optimal parameter more difficult to be identifiable.

In contrast, non-parametric test methods such as the Kolmogorov-Smirnov test (K-S) do not require explicitly calculating of the summary statistics of data but comparing two time series at each time step with cumulative distribution probability, which in principle can avoid the loss of high-frequency data information suffered by existing works. In 2022, the work \cite{21} used K-S for calibration. However, this work did not explore the problem of high-frequency data information loss of K-S, nor was it verified on real market data. Is it feasible to calibrate high-frequency data using K-S as the objective function? Are there any technical challenges? This work aims to answer these questions.

This paper first visualizes the parameter space of the K-S test objective function under high-frequency calibration and finds that the optimal solution (non-identifiable) area in the high-frequency calibration parameter space of K-S is greatly reduced compared with the calibration of low-frequency data. On one hand, this shows that K-S has better parameter identifiability. This suggests that K-S does not cause too much data information loss, making it a proper objective function for high-frequency data calibration. To this end, this paper designs a set of experiments to compare the K-S and traditional MSM as high-frequency data calibration objective functions, confirming that K-S leads to a higher fidelity of simulation. On the other hand, many discontinuous local optimal areas are observed with the reduction of the non-identifiability. The multi-modal parameter space makes the optimization even difficult. In other words, it is feasible to use K-S as a high-frequency calibration objective function, but at the same time, a more powerful optimization algorithm is required.

In this paper, the Negatively Correlated Search (NCS) algorithm \cite{22} is used to address the multi-modal characteristics of the high-frequency calibration problem. NCS constructs multiple randomized iterative search processes based on the Gaussian distribution.  At each iteration, the search processes are guided by the explicitly modeled expected objective function values of all search processes and the distributional distance between pairwise search processes. By optimizing these two items simultaneously, multiple search processes can collaboratively and concurrently explore different yet promising regions in the parameter space, which is particularly suitable for multi-modal optimization problems \cite{23,24,25}. Among them, balancing the objective function value and the diversity between search processes is the key to controlling the collaboration of multiple search processes. Note that, these two terms are often at different scales in this calibration problem. Hence, traditional balancing strategies like weighted average do not work well. To address this issue, this paper designs a novel balancing strategy called the adaptive stochastic ranking to improve the capability of calibrating the FMSs with high-frequency data.

This paper takes the PGPS (Preis-Golke-Paul-Schneider) liquidity provider-taker model as the studied FMS \cite{7}, which has attracted extensive research attention in recent years. Based on PGPS, this paper reports the high-fidelity calibration results at second-level for both synthetic data and real data for the first time. Compared with the commonly adopted algorithms, the improved NCS increases the simulation fidelity by $16.3\%$ and $36.0\%$ on synthetic data and by $5.6\%$ and $16.6\%$ on real data, respectively.

The remainder of this paper is organized as follows. Section 2 provides the background about the financial markets and the employed PGPS simulator. Section 3 analyzes the characteristics and difficulties of the K-S objective function under high-frequency data calibration in detail. Section 4 introduces the improved NCS. Section 5 experimentally verifies the effectiveness of K-S as a high-fidelity objective function for calibrating high-frequency data. Section 6 shows the second-level calibration performance of the proposed NCS variant on both synthetic data and real data. Section 7 concludes this paper.

\section{Background}

\subsection{Order Matching and Market State Data}

\noindent The regular trading hours of market at a day is divided into two stages: call auction and continuous auction. Taking the Chinese A-share market as an example, 9:15 to 9:25 is the opening call auction time, 9:30 to 11:30 and 13:00 to 14:57 is the continuous auction time, and 14:57 to 15:00 is the closing call auction time.

\begin{figure}[htbp]
\centering
\includegraphics[scale=0.5]{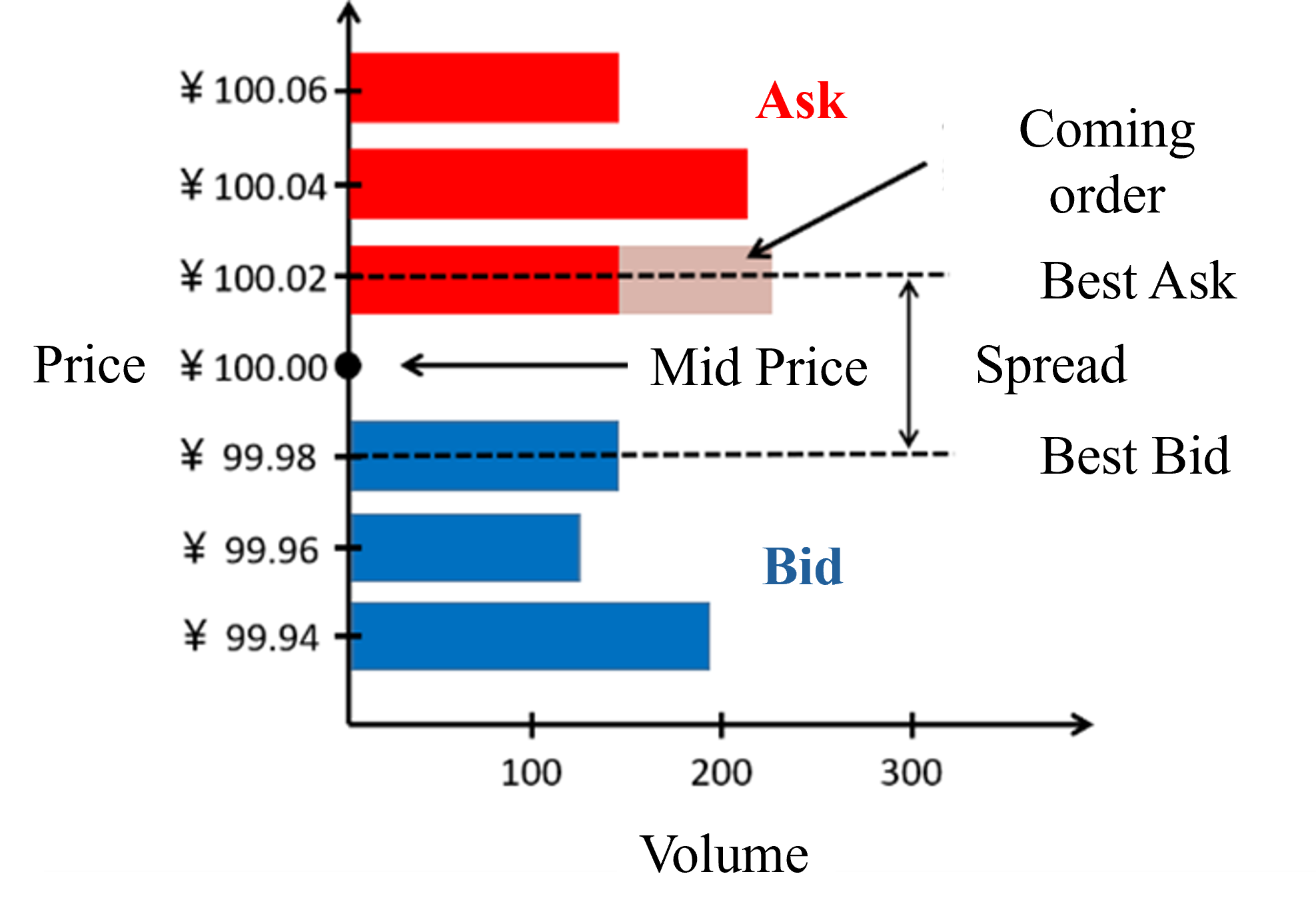}
\caption{The Diagram of the Limit Order Book.}
\label{figure1}
\end{figure}

During the continuous auction period, each trader in the market can submit three types of orders to the exchange: limit orders, market orders and cancel orders. A limit order is an order that sets a price and volume, and specifies the direction (ask for selling or bid for buying). A market order is an order that is executed instantly at the current best price with specified direction and volume. A cancel order is to delete an order that has not been fully traded. The exchange receives orders submitted by all traders and maintains a limit order book (LOB) to record all orders that are not fully traded \cite{26,27}. As shown in Fig.\ref{figure1}, the price most likely to be executed in both directions is called the first-level bid price/first-level ask price. The average of the first-level bid price and the first-level ask price is called the mid-price. Previous researchers mainly used the mid-price as market state data $\widehat{s}_t$. The remaining levels of prices are sorted from the mid-price to both sides. The price of a market order is automatically set to the price of first-level at the opposite direction.

The major exchanges worldwide all adopt a continuous double auction (CDA) mechanism \cite{26} to match the new incoming orders one by one. Let us take a bid order as an example. Asking orders follow the same principle but operate in the opposite direction. If it is a limit order, CDA will match the new bid order in a "price-first-time-second" priority in the selling direction of the LOB. That is, the bid order tries to match the ask orders from the lowest price levels to the highest. Once it matches successfully, the earliest coming selling orders at that price level will be deleted until the volume of the new bid order is fully traded. If it does not match any price level, i.e., the price of the bid order is lower than the lowest (first-level) ask price), it will be inserted into the corresponding level of at the bid direction of the LOB according to its bid price. If it is a market order, the bid order price will be automatically set to the first-level ask price and then matched according to the above "price-first-time-second" priority.

LOB updates in real time during the continuous bidding phase, which is determined by the frequency of incoming orders. When the exchange publishes the state data to the traders, the real-time LOB will be processed at different frequencies to meet the needs of different data consumers. Generally, $\{\widehat{s}_t\}_{t=1}^T$ with a frequency below 1 second is considered as high-frequency data, which can be used to serve high-frequency trading such as intraday trading. Data with a frequency above daily is considered low-frequency.

Distinct from the tick-level matching mechanism of continuous auction, the call auction is a one-shot matching of all bid and ask orders submitted within the call auction period, aiming to maximize the trading volume among those orders. Due to its simplicity, almost all literature takes its matching result as the initial LOB of the continuous auction period, while the latter is the major duration that researchers strive to simulate.  

\subsection{The Simulator $M(w)$}

\noindent The financial market is a complex system including two major components: an exchange and many traders. To better simulate the financial market, the ABM techniques normally model the exchange (the order matching mechanism) and the traders (the orders submission strategies) separately. For the order matching mechanism, the above mentioned CDA is essentially a deterministic rule-based calculation process. This paper adopts the well-established MAXE environment \cite{28}. For the orders submission strategies, there is no accurate mathematical models due to the uncertainty of human traders. Researchers have proposed various agent models to approximate human traders by introducing knowledge and findings from economics and finance. This efforts indeed successfully simulate various trading strategies observed from the real market \cite{6}. For example, the work \cite{7} proposed the Preis-Golke-Paul-Schneider(PGPS) model based on the theory of market liquidity provider and liquidity taker. PGPS model is capable of simulating the trading behaviors of submitting limit orders, market orders, and order cancellations. The work \cite{8} proposed the fundamentalist-chartist model to simulate the trading behaviors which trade based on either fundamental value or K-line. This work adopts the PGPS model for calibration, as it has become very popular in recent years.

\begin{figure}[htbp]
\centering
\includegraphics[scale=0.3]{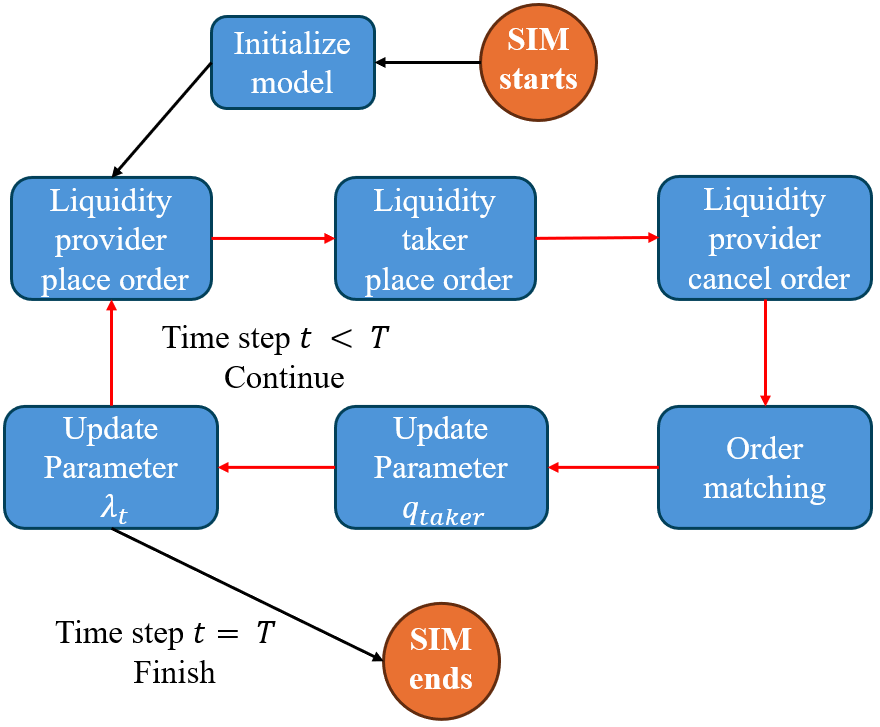}
\caption{The Workflow of the PGPS Simulator.}
\label{figure2}
\end{figure}

The PGPS model abstracts all traders in the market into two categories of trading agents: $N_A$ liquidity providers and $N_A$ liquidity takers. The overall simulation procedure is shown in Fig.\ref{figure2}. The red loop will execute $T$ rounds. Each liquidity provider submits a limit order at each time step $t$ with a probability of $\alpha$, and the default order volume is $1$. The probability of each order being a buy/sell order is set to $0.5$. Each liquidity taker submits a market order at each time step $t$ with a probability of $\mu$, and the default order volume is set to $1$. The probability of each order being a buy order or a sell order is $q_{taker}(t)$ or $1-q_{taker}(t)$, respectively. $q_{taker}(t)$ follows a mean reversion random walk process with a mean of $0.5$. The probability of approaching the mean is $0.5+\left|q_{taker}(t)-0.5\right|$ and the approaching step size is $\pm\Delta_s$. In addition, each liquidity taker will cancel an untraded limit order at each time step $t$ with a probability of $\delta$. The price of a limit order, either bid or ask, is calculated as Eq.\eqref{eq2} and Eq.\eqref{eq3}, respectively

\begin{equation}
bid\; price=p_a(t)-1-\lfloor-\lambda(t)\log u\ \rfloor,
\label{eq2}
\end{equation}
\begin{equation}
ask\; price=p_b(t)+1-\lfloor-\lambda(t)\log u\ \rfloor,
\label{eq3}
\end{equation}

\noindent where $p_a(t)$ is the current first-level ask price, $p_b(t)$ is the current first-level bid price, and $u\sim U(0,1)$ is a uniform random number. $\lambda(t)$ is a time-dependent parameter indicating the depth of order placement and is calculated as Eq.\eqref{eq4}.

\begin{equation}
\lambda(t)=\lambda_0\ (1+\frac{|q_{taker}(t)-0.5|}{\sqrt{\langle(q_{taker}(t)-0.5)^2 \rangle}}C_{\lambda}).
\label{eq4}
\end{equation}

\noindent Here, $\langle(q_{taker}(t)-0.5)^2 \rangle$ is the average value of $(q_{taker}(t)-0.5)^2$ by taking $10^5$ Monte Carlo sampling at every time-step during the simulation loop.

In the PGPS model, the key parameters are $w=[\alpha, \mu, \delta, \Delta_s, \lambda_0, C_{\lambda}]$. Among them, $[\alpha, \delta, \lambda_0, C_{\lambda}]$ are shared by all liquidity providers and $[\mu,\Delta_s]$ are shared by all liquidity takers. In other words, there are 6 model parameters calibrated in this paper in total. Followed by the work \cite{19}, the number of two types of agents is set to $N_A=125$. The calibration algorithm only needs to calibrate these 6 hyperparameters. The remaining parameters are adaptively updated by the above formulation indicating the trading strategies of the agents.

\section{The Properties of Calibrating High-frequency data with K-S}

\noindent In this paper, we consider K-S test as the metric $D$. By adopting K-S, Eq.\eqref{eq1} can be further written as Eq.\eqref{eq5}.

\begin{equation}
\mathop{min}\limits_w\mathop{sup}\limits_x{|\frac{1}{T}\sum_{t=1}^{T}{I(\widehat{s}_t\le x)}-\frac{1}{T}\sum_{t=1}^{T}{I(s_t\le x)}|}.
\label{eq5}
\end{equation}

\noindent$I(\cdot)$ is an indicator function, which outputs 1 if the input is true, otherwise it outputs 0. For simplicity, the value of $x$ is set to the value of each element in $\{{s_t}\}_{t=1}^T$ and $\{{\widehat{s}_t}\}_{t=1}^T$, which is used to calculate the cumulative distribution of $\{{s}_t\}_{t=1}^T$ and $\{{\widehat{s}_t}\}_{t=1}^T$.

With the simulator described in Section 2.2, we first visualize the parameter space to analyze the properties of this K-S calibration objective function. Specifically, given the objective function $D(\{{\widehat{s}_t}\}_{t=1}^T, M(w)=\{{s_t}\}_{t=1}^T)$ and input the target data $\{{\widehat{s}_t}\}_{t=1}^T$, we can choose any pair of the 6 parameters in $w$. Then, for the chosen pair of parameters, we uniformly sample 100 values for each parameter in the parameter space and depict the landscape of the two-dimensional parameter space with a total of $100\times 100=10000$ grid samples. The brightness of each area indicates its objective function value. The brighter the areas are, the smaller objective function values they have. For better visualization, only the top-2000 parameters with the smallest objective function values among the 10000 uniform samples are depicted. The red dot represents the optimal parameter who generates the target data $\{{\widehat{s}_t}\}_{t=1}^T$ with simulator $M$.

\begin{figure}[t]
\centering
\includegraphics[scale=0.9]{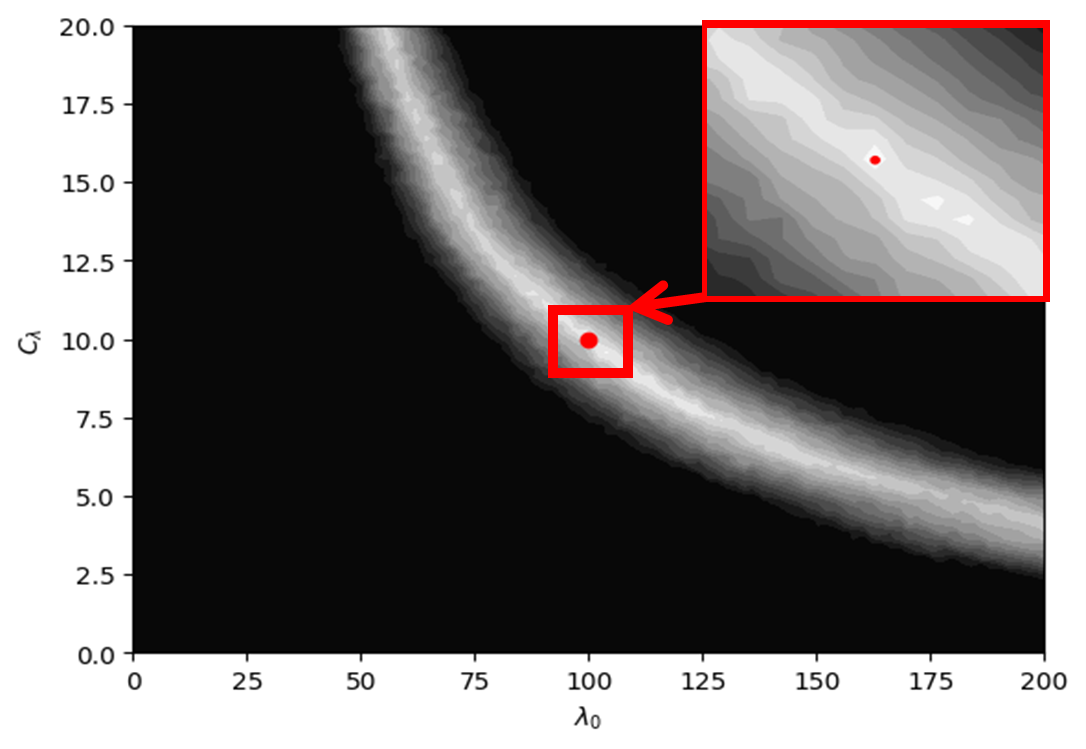}
\caption{Two-dimensional parameter space of K-S with target data of 1-second level frequency.}
\label{figure3}
\end{figure}

\begin{figure}[t]
\centering
\includegraphics[scale=0.9]{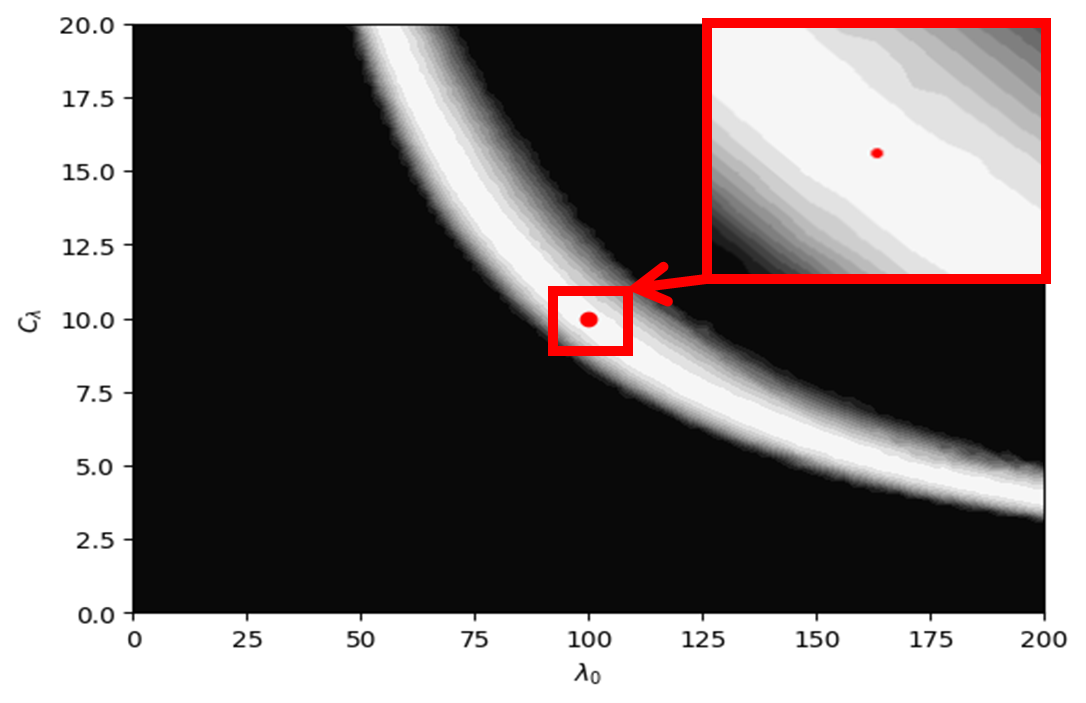}
\caption{Two-dimensional parameter space of MSM with target data of 1-second level frequency.}
\label{figure4}
\end{figure}

First, the differences of the landscape of two objective functions $D$ under the same target data $\{{\widehat{s}_t}\}_{t=1}^T$ are compared. The parameter pair $[\lambda_0, C_{\lambda}]$ is chosen. The two-dimensional parameter space of the K-S objective function and the most commonly used MSM \cite{5} objective function are visualized in Fig.\ref{figure3} and Fig.\ref{figure4}, respectively. It shows that the parameter space of K-S is more identifiable than MSM. The brightest area in the upper right corner is significantly smaller. This means that the K-S objective function has less information loss when calculating the difference of two data sequences. The same phenomenon exists in other parameter pairs. But due to the multi-modal characteristics of the landscape, the visualization effect is not as intuitive as the $[\lambda_0, C_{\lambda}]$ parameter pair.

\begin{figure}[htbp]
\centering
\includegraphics[scale=0.9]{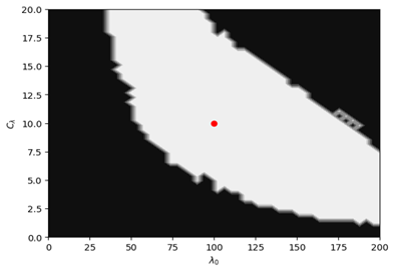}
\caption{Two-dimensional parameter space of K-S with target data of 30-minute level frequency.}
\label{figure5}
\end{figure}

Secondly, the frequency of the target data $\{{\widehat{s}_t}\}_{t=1}^T$ is decreased from 1-second level as for depicting Fig.\ref{figure3} and Fig.\ref{figure4} to 30 minutes level to see how sensitive K-S is to different frequency data. The same two-dimensional parameter space (i.e., $[\lambda_0, C_{\lambda}]$) with the lowered resolution (i.e., 30 minutes level) is visualized in Fig.\ref{figure5}. It can be seen that the objective function values of the top-2000 samples are with the same brightness. Compared to Fig.\ref{figure3}, it shows that with the increase in the data frequency, the identifiability of the parameter space of the K-S objective function is largely improved. This suggests that K-S is a high-fidelity objective function for calibrating high-frequency data as it can preserve more detailed high-frequency information in the landscape (e.g., in Fig.\ref{figure3}).

Specifically, the calculation process of both K-S and MSM includes two loops. The inner loop is the characterization of the distributional difference between the two sets of data, and the outer loop is the aggregation index of the distributional difference. The advantages of the K-S objective function can be explained at both loops: 1) K-S uses the distance of cumulative distribution function value to measure the difference between the two sets of data, and is less affected by the increase of data dimensionality. Comparatively, if the summary statistics used by the MSM are not sufficient statistics, it will cause different degrees of the information loss of the target data. Since the distributions of financial data usually do not belong to the exponential family, its summary statistics are normally not sufficient statistics. 2) In the the outer loop of Eq.\eqref{eq5}, the $max$ aggregation function of K-S can provide higher identifiability than the weighted average of MSM. Compared with the average function, the $max$ function is less likely to weaken the difference measurement of the two sets of data with the increase of data dimensionality, which is more effective for high-frequency data which involves longer (higher dimensional) time series to be calibrated.

\begin{figure}[t]
\centering
\includegraphics[scale=0.275]{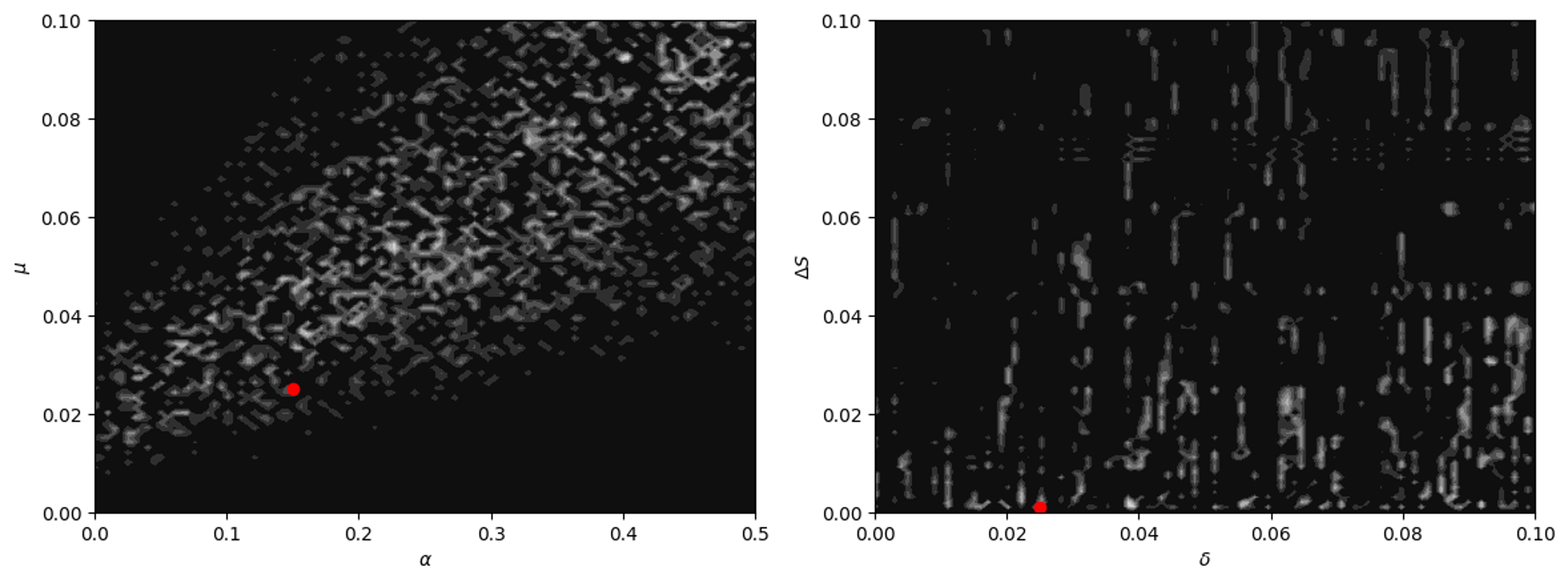}
\caption{Two-dimensional parameter space of K-S Objective Function in either $[\alpha, \mu]$ (left) or $[\delta, \Delta_s]$ (right).}
\label{figure6}
\end{figure}

Finally, the two-dimensional parameter space with respect to other pairs of parameters of the K-S objective function is depicted with the 1-second level data frequency. In particular, we took the two groups of parameter combinations $[\alpha, \mu]$ and $[\delta, \Delta_s]$ as examples and plotted them in Fig.\ref{figure6}, respectively. It is observed that these two dimensional parameter space appears to multi-modal where many bright spots (i.e., local optima) scatter over the landscape, and these optimal areas are topologically strongly discontinuous. This makes the K-S high-frequency calibration problem, though more identifiable, but more difficult to find the global optimum. If the optimization algorithm lacks global exploration capabilities, it is easy to fall into the local optimal areas. In other words, solving the high-frequency calibration problem requires a more powerful multi-modal optimization algorithm.

\section{The Improved NCS}

\subsection{The NCS Framework}

\noindent Since the objective function of Eq.\eqref{eq5} in this paper is not differentiable, the calibration has to resort to the black-box optimization algorithms which are basically randomized iterative search processes. The parameter space of high-frequency data calibration problem has been analyzed to be multi-modal. Traditional black-box optimization algorithms involving single search process are found difficult to solve multi-modal problems. The difficulty lies in that a single search process often converges to one of the many local optima. Therefore, an optimization algorithm that can effectively locate multiple optima can be promising as it enhances the probability that one of the located optima shows to be the global optimum. 

NCS is a black-box optimization algorithm that formally models the target of "searching for multiple different optima" as a maximization problem \cite{23,22}, and directly derives the search method by solving such maximization problem. Specifically, the model consists of multiple Gaussian distribution based search processes \cite{30,31} and the derived NCS basically acts to coordinate those search processes to converge to different optima in the parameter space. 

Assume that there are $\lambda$ search processes, each of which is modeled as a multivariate Gaussian distribution $p(\theta)=\mathcal{N}(m,\Sigma)$ in the parameter space. Searching for new candidate parameters is equivalent to randomly sampling from the Gaussian distribution and the parameters closer to the mean of the distribution has a higher probability of being sampled. The Gaussian distribution is updated as the optimization process iterates. First, one offspring is generated by $\theta^{\prime}\sim p(\theta)$. Then the better one between the parent distribution $p(\theta)$ and the offspring distribution $p(\theta^{\prime})$ is selected to enter the next iteration. That is, keep $\theta$ unchanged or update it to $\theta^{\prime}$. The selection process is determined by two factors: 1) The expected objective function of the distributions. 2) The distances between distributions. For each $i$-th Gaussian distribution,

\begin{enumerate}
\item[1)] The former emphasizes the exploitation of the optimization and is described by the objective function value of all $N$ candidate parameters sampled from the Gaussian distribution, i.e.,

\begin{equation}
F(\theta_i)=\sum_{k=1}^{N}{f(w_i^k)p(w_i^k|\theta_i)}
\label{eq6}
\end{equation}

For each parameter $w_i^k$, $f(w_i^k)$ is calculated as: first, the simulated data sequence is generated by $M(w_i^k)=\{{s_t}\}_{t=1}^T$, then by applying the K-S test we have $f(w_i^k)=\mathop{sup}\limits_x{|\frac{1}{T}\sum_{t=1}^{T}{I(\widehat{s}_t\le x)}-\frac{1}{T}\sum_{t=1}^{T}{I(s_t\le x)}|}$ based on Eq.\eqref{eq5}. Minimizing $F(\theta_i)$ can theoretically make the $i^{th}$ search process converge to a local optimum.

\item[2)]The latter highlights the exploration by enhancing the diversity of the search processes. It measures the how likely the area may be sampled in the next iteration by calculating the distributional distance between pairwise search processes (e.g., the Bhattacharyya distance \cite{22,23}), i.e.,

\begin{equation}
\normalsize
D(\theta_i)=-\sum_{k=1}^{\lambda}\log{(}\sum_{k=1}^{N}\sqrt{p(w_i^k|\theta_i)p(w_i^k|\theta_j)})
\label{eq7}
\end{equation}

Maximizing $D(\theta_i)$ can theoretically make the $i^{th}$ search process converge to the region in the parameter space that does not overlap with other search processes.
\end{enumerate}

Hence, the trade-off between $F$ and $D$ determines the bias of the search process towards either exploration or exploitation, which strongly influences the selection between the parent and offspring distributions to the next iteration, and further impacts on the performance of NCS.

\subsection{The Adaptive Stochastic Ranking}

\noindent The paper \cite{23} proposed that when $F(\theta_i^{\prime})/D(\theta_i^{\prime})<\mathcal{N}(1,\xi)$, the offspring distribution $p(\theta^{\prime})$ should be selected for sampling in the next iteration. Otherwise, the parent distribution $p(\theta)$ should be selected. $\xi$ decays linearly from 0.1 to 0 as the search process iterates. And $F(\theta_i^{\prime})$ and $D(\theta_i^{\prime})$ are the expected objective function of the $i^{th}$ offspring distribution $\theta_i^{\prime}$ and the diversity between $\theta_i^{\prime}$ with other distributions, respectively. This method relies on two settings: 1) The dynamics of $F(\theta_i^{\prime})/D(\theta_i^{\prime})$ conforms to $\mathcal{N}(1,\xi)$. 2) The dynamics of $\xi$ corresponds to the total number of iterations. These two settings vary over problems. Hence, a good performance in the original paper does not mean that it will also be good in the calibration problem considered in this paper. In addition, the method does not explicitly consider the parent's $F(\theta_i)$ and $D(\theta_i)$, which may not be accurate enough.

To achieve a better balance between $F$ and $D$, this paper considers $F(\theta_i)$, $F(\theta_i^{\prime})$, $D(\theta_i)$ and $D(\theta_i^{\prime})$ at the same time. This naturally leads to four cases:

\begin{enumerate}
\centering
\item[1)] $F(\theta_i^\prime)<F(\theta_i)\land D(\theta_i^\prime)>D(\theta_i)$,
\item[2)] $F(\theta_i^\prime)>F(\theta_i)\land D(\theta_i^\prime)>D(\theta_i)$,
\item[3)] $F(\theta_i^\prime)<F(\theta_i)\land D(\theta_i^\prime)<D(\theta_i)$,
\item[4)] $F(\theta_i^\prime)>F(\theta_i)\land D(\theta_i^\prime)<D(\theta_i)$.
\end{enumerate}

Note that, $F$ is a minimization problem and $D$ is a maximization problem. Therefore, for case 1), the offspring is better in both $F$ and $D$, so the offspring should be selected. Similarly, for case 2), the parent should be selected. For cases 3) and case 4), it is difficult to determine whether to choose the offspring or the parent. However, if the importance of $F$ and $D$ is determined, it should be intuitive based on this priority. How to establish the priority of $F$ and $D$ so that exploration and exploitation can be balanced better? This paper proposes a novel mechanism called the adaptive stochastic ranking as follows.

For case 3), considering the goal of the calibration problem is to obtain optimal parameters with respect to $F$, it is intuitive that the priority of $F$ over $D$ should increase with the number of iterations. To this end, a deterministic priority trade-off strategy is designed for case 3). Assume that $D$ takes precedence over $F$ with a certain probability $\beta$ and $F$ takes precedence over $D$ with a probability of $1-\beta$. The probability $\beta$ decays at a certain rate with the number of iterations, i.e., $\beta=0.7-\frac{0.4G}{G_{max}}$. $G$ indicates the current number of iterations and $G_{max}$ is the total number of iterations. In other words, $D$ may have a higher priority in the early stage of calibration, while $F$ is more preferred in the later stage of calibration.

For case 4), it cannot be considered as the symmetric situation of case 3), as it contains more complex distributions of population. Therefore, it cannot be regulated based on the above deterministic priority trade-off strategy. This paper proposes to take an adaptive strategy to obtain more feedback from the search processes to trade-off the priority of $F$ and $D$. For this purpose, we design a time-varying threshold parameter $\epsilon_t$ and an individual update rate parameter $\varphi_t$ (the proportion of the number of offspring that successfully replaces the parent). Intuitively, when the individual update rate is less than $\epsilon_{t}$, the exploitation of the current area should be increased. That is, in the next iteration, $D$ takes precedence over $F$. It should be emphasized that $\epsilon_{t}$ is not fixed. As shown in Eq.\eqref{eq8}, if the individual meets the condition that the update rate $\varphi_{t-1}$ is greater than $\epsilon_{t-1}$ in the last iteration, $\epsilon_t$ should be updated as $\epsilon_{t-1}$ multiplying a fixed step $\rho$. If $\varphi_{t-1}$ does not meet the condition, $\epsilon_t$ is reset to the initial value $\epsilon$. This ensures that the chances of meeting the conditions increases exponentially, avoiding degeneration to population diversity caused by over-exploitation.

\begin{equation}
\epsilon_t=\left\{
\begin{aligned}
&\epsilon_{t-1} \times \rho & if\; \varphi_{t-1} > \epsilon_{t-1}\\
&\epsilon & if\; \varphi_{t-1}\le \epsilon_{t-1}
\end{aligned}
\right .
\label{eq8}
\end{equation}

\subsection{The Pseudo-Code of the Improved NCS}

\noindent Based on the above discussions, the workflow of the improved NCS can be described as follows.

\begin{breakablealgorithm}
    \caption{Improved NCS}
    \label{NCS}
    \renewcommand{\algorithmicrequire}{\textbf{Input:}}
    \renewcommand{\algorithmicensure}{\textbf{Output:}}
    
    \begin{algorithmic}[1]
        \REQUIRE $\lambda$, $N$  
        \ENSURE $w^*$    
        
        \STATE  Initialize $\lambda$ search processes $\{p(\theta_i)\}_{i=1}^\lambda$, each $i$-th search process follows a Gaussian distribution $\mathcal{N}(m_i,\Sigma_i)$.
        
        \FOR{$i = 1 : \lambda$}
            \STATE Sample $N$ candidates $\{w_i^k\}_{k=1}^N$ from $p(\theta_i)$ and evaluate them as $\{f(w_i^k)_{k=1}^N\}$.
            \STATE Calculate $F(\theta_i)$ and $D(\theta_i)$ with Eqs.\eqref{eq6}-\eqref{eq7}.
        \ENDFOR

        \FOR{$G = 1 : G_{max}$}
            \FOR{$i = 1 : \lambda$}
                \STATE Generate offspring distribution $\theta_i^\prime\sim p(\theta_i)$.
                \STATE Sample $\{w\prime_i^{k}\}_{k=1}^N$ from $p(\theta_i^\prime)$ and evaluate them as $f(w\prime_i^{k})$.
                \STATE Calculate $F(\theta_i^\prime)$ and $D(\theta_i^\prime)$ with Eqs.\eqref{eq6}-\eqref{eq7}.
                \IF{$F(\theta_i^\prime)<F(\theta_i)\land D(\theta_i^\prime)>D(\theta_i)$}
                    \STATE $\theta_i=\theta_i^\prime$.
                \ELSIF{$F(\theta_i^\prime)>F(\theta_i)\land D(\theta_i^\prime)>D(\theta_i)$}
                    \STATE $\theta_i=\theta_i^\prime$ with probability $\beta=0.7-\frac{0.4G}{G_{max}}$.
                \ELSIF{$F(\theta_i^\prime)<F(\theta_i)\land D(\theta_i^\prime)<D(\theta_i)$}
                    \IF{$\varphi_t > \epsilon_t$}
                        \STATE $\theta_i=\theta_i^\prime$.
                    \ENDIF
                \ENDIF
            \ENDFOR
            \STATE Update $\epsilon_t$ according to Eq.\eqref{eq8}.
            \STATE Update $w^*$ as the best parameter in terms of $f$.
        \ENDFOR
    \end{algorithmic}
\end{breakablealgorithm}

The algorithm first initializes $\lambda$ search processes following the Gaussian distribution in step 1. $m_i$ is randomly sampled from the parameter space and $\Sigma_i$ is a $6\times6$ diagonal matrix where the elements are set to $\frac{1}{\lambda}$ the interval between upper bound and lower bound of the parameter space. Steps 2 to 5 evaluate the first generation of each search processes with Eqs.\eqref{eq6}-\eqref{eq7}. Steps 6 to 23 describe the process of iterative optimization which is the main part of the algorithm. In each iteration, the offsprings are generated from the last generation, which are also evaluated by the Eqs.\eqref{eq6}-\eqref{eq7}. Then, with the rules described in Sections 4.1 and 4.2, the generations and algorithm parameters are continuously updated until the current number of iterations $G$ equals the given number $G_{max}$. Finally, the parameter $w^*$ with the best objective function $f(w^*)$ is output as the result.

\section{Simulations on the K-S objective function}

\noindent This section empirically demonstrates the effectiveness of using K-S test as the objective function for high-frequency data calibration.

\subsection{The Descriptions of the Target Data $\{{\widehat{s}_t}\}_{t=1}^T$}

\noindent 10 target data instances are synthesized based on the MAXE simulation environment and the PGPS model $M(w)$ by setting 10 different target parameters, respectively. All the target data instances are the mid-price data, which is the average of the best bid price and the best ask price in the LOB. Each of the 10 (6-dimensional) parameters is randomly sampled within the pre-defined search range. The search range of each parameter are directly borrowed from the literature \cite{19}, as shown in Table \ref{table1}. The synthetic data are all 1-second level data generated by the simulator $M(w)$ with a simulation length of an hour, i.e., $T=3600$ simulation time steps in total. Generally, the larger $T$ is, the more dynamics the data may involve and it is more difficult to calibrate the model. In comparison, the literature \cite{19} uses 1-minute level data for one week to calibrate the PGPS model, which has only 2300 simulation time steps. In other words, the synthetic data adopted in this paper is not trivial to calibrate. To eliminate the bias of randomness, the 10 sets of target data are verified to involve sufficient diversity in the statistical properties like trend and fluctuation range, so as to represent different market environments.

\begin{table}[H]
\caption{\textbf{The Range of Parameters for Synthesizing Data}}
\label{table1}
\vspace{8pt}
\centering
\begin{tabular}{ccc}
\toprule
\textbf{Parameter} & \textbf{Range} & \textbf{Description}\\
\midrule
$\alpha$ & $[0.05,0.20]$ & Prob. of limit order\\
$\mu$ & $[0.00,0.05]$ & Prob. of market order \\
$\delta$ & $[0.00,0.05]$ & Prob. of order canceling\\
$\Delta_s$ & $[0.00,0.005]$ & Mean reversion step \\
$\lambda_0$ & $[50.00,300.00]$ & Initial depth of LOB \\
$C_{\lambda}$ & $[1.00,50.00]$ & Step size for varying LOB \\
\bottomrule
\end{tabular}
\end{table}

\subsection{Experimental Settings}

\noindent The experiment takes the most commonly used MSM metric in the FMS literature as the compared calibration objective function. MSM basically measures the mean square error of the moments between the target data and the simulated data, which is generated by the calibrated parameters. This paper selects the four commonly seen moments such as mean, standard deviation (std), skewness, and kurtosis of the 1-second level mid-price \cite{21}, as shown in Table \ref{table2}. For each target data instance, K-S and MSM are used as the objective function for calibration, respectively. The best-found parameter using Algorithm 1 is applied into the simulator to generate simulated data. Then the difference between each simulated data and its corresponding target data is evaluated using MSM and K-S as performance indicators. For fairness, K-S and MSM are used as performance evaluation indicators to show the consistency of K-S and MSM. For each calibration, the iteration number is set to 10000, which means that a total number of 10000 parameter samplings and evaluations are performed for each target data.

\begin{table}[tbhp]
\caption{\textbf{The Mean, Standard Deviation, Skewness and Kurtosis of the Mid-price}}
\label{table2}
\vspace{8pt}
\centering
\begin{tabular}{cc}
\toprule
\textbf{Selected Moments} & \textbf{Formulaic Definition} \\
\midrule
Mean & $\frac{1}{T}\sum_{t=1}^{T}s_t$ \\
Std & $\sqrt{\frac{\sum_{t=1}^{T}(s_t-mean)^2}{T-1}}$ \\
Skewness & $\frac{1}{T-1}\sum_{t=1}^{T}(\frac{s_t-mean}{std})^3$ \\
Kurtosis & $\frac{1}{T-1}\sum_{t=1}^{T}(\frac{s_t-mean}{std})^4$ \\
\bottomrule
\end{tabular}
\end{table}

\subsection{Results and Analysis}

\noindent As shown in Table \ref{table3}, the calibration results using K-S as the calibration objective function are closer to the target data than the calibration results using MSM as the objective function. The calibration of K-S results in smaller distances to the target data under both K-S and MSM indicators. Comparatively, by using MSM as the performance indicator, the calibration results of K-S are significantly better than that obtained by MSM on almost all the instances. For the only inferior case (instance 5), the MSM indicator value of the K-S calibration result is also very close to the MSM calibration result. This not only verifies the feasibility of K-S in high-frequency data calibration but also shows that the high fidelity of K-S can somehow enable the calibration process to find a better parameter faster.

\section{Simulations on the Improved NCS}

\noindent This section mainly answers the following three questions through experimental simulations, to verify the calibration capability on high-frequency data of the proposed improved NCS:

\begin{table}[H]
\caption{\textbf{K-S as Calibration Objective is Better than MSM Whenever the Performance Indicator is K-S or MSM}}
\label{table3}
\vspace{8pt}
\centering
\begin{tabular}{ccccc}
\toprule
\textbf{Objective Function} & \multicolumn{2}{c}{\textbf{K-S}} & \multicolumn{2}{c}{\textbf{MSM}} \\
\midrule
\textbf{Performance Indicator} & K-S & MSM & K-S & MSM \\
\midrule
Instance 1 & \textbf{0.026} & \textbf{0.86} & 0.081 & 11.17 \\
Instance 2 & \textbf{0.058} & \textbf{59.33} & 0.108 & 92.10 \\
Instance 3 & \textbf{0.035} & \textbf{3.09} & 0.274 & 29.58 \\
Instance 4 & \textbf{0.036} & \textbf{8.18} & 0.111 & 29.97 \\
Instance 5 & \textbf{0.029} & 2.31 & 0.063 & \textbf{2.12} \\
Instance 6 & \textbf{0.031} & \textbf{12.00} & 0.334 & 64.86 \\
Instance 7 & \textbf{0.052} & \textbf{12.55} & 0.192 & 37.47 \\
Instance 8 & \textbf{0.040} & \textbf{0.43} & 0.296 & 10.80 \\
Instance 9 & \textbf{0.070} & \textbf{27.22} & 0.791 & 189.58 \\
Instance 10 & \textbf{0.044} & \textbf{2.69} & 0.078 & 3.96 \\
\bottomrule
\end{tabular}
\end{table}

\begin{enumerate}
\item[1)] Can NCS obtain parameters with better objective function values on this multi-modal optimization problem?
\item[2)] Is the simulated data generated from parameters calibrated by NCS more consistent with the target synthetic data?
\item[3)] Can real historical 1-second level data be calibrated with high fidelity?
\end{enumerate}

\subsection{The Description of Target Data $\{{\widehat{s}_t}\}_{t=1}^T$}

\noindent There are two sets of data instances considered in this section. The first set of target data is synthetic data generated by the simulator, whose settings are consistent with Section 5.1. The second set of target data is the real historical data. The paper chooses the 1-second level mid-price data of the US Total Market Index (CRSP US Total Market Index) from 9:30 am to 10:30 am on June 28, 2023, i.e., 3600 time steps in total. For the calibration on real historical data, the ground-truth optimal parameter cannot be known in advance. Therefore, one additional difficulty of calibrating real data is that the search range of each parameter cannot be determined a priori, resulting in a potentially larger parameter space and thus more difficulty in search.

\subsection{Algorithm Settings}

\noindent Recently, the work \cite{21} proposed the Trust Region Bayesian Optimization (TuRBO) \cite{29,30} as a calibration algorithm with the K-S objective function and achieves promising results on synthetic data. TuRBO considers the insufficient capability of global Gaussian surrogate model on capturing the heterogeneous structure of the entire parameter space as the main difficulty of Bayesian optimization. Therefore, it proposes to maintain multiple local surrogate models and allocate search resources globally among these local surrogate models. Approximate Bayesian Computation (ABC) \cite{31} is another widely studied method for FMS calibration \cite{32}. Approximate Bayesian computation (ABC) is a numerical method for Bayesian inference, which applies to processing complex models where the likelihood function is difficult to calculate. It iteratively randomly samples parameters from the parameter space based on the proposal distribution and generates simulation data based on the model. Then it selects candidate parameters based on the discrepancy between the simulation data and the target data. After that, ABC methods update the proposal distribution according to the selected parameters and avoid the difficulty of directly calculating the likelihood function. This paper takes these two algorithms as compared algorithms to verify the improved NCS on the calibration of high-frequency data under the K-S objective function.

The parameter setting of the improved NCS is consistent with the original suggestions \cite{22}, which mainly includes two parameters $\lambda=10$ and $N=1$.  According to the work \cite{21}, TuRBO samples and evaluates 10 candidate parameters in each iterative search to maintain the local surrogate models. The rest parameter settings are directly borrowed from the original paper \cite{21}. The number of samples in every generation of ABC is set to 10 to keep the samples per iteration of three algorithms the same. The rest parameter settings of ABC follow the paper \cite{33}.

To eliminate the bias of randomness, for the two sets of data instances, NCS, TuRBO, and ABC run 10 times with different random numbers. Each calibration process includes 10000 parameter samplings and simulation evaluations. Since NCS, TuRBO, and ABC all sample and evaluate 10 individuals in each iteration, all three algorithms will iterate 1000 times in each run. The searched parameter with the minimum K-S objective function value is the output of each run. The performance of the 10 independent runs is averaged as the perforamnce of the algorithm. The one out of ten parameters with the smallest K-S objective values is selected as the best parameter, and its resultant simulated data is regarded as the best simulated data.

\begin{table}[tbh]
\caption{\textbf{Performance on Different Indicators of TuRBO, ABC and NCS on 10 Synthetic Data with 10 Repeated Runs}}
\label{table4}
\vspace{8pt}
\centering
\begin{tabular}{ccccc}
\toprule
\textbf{Indicators} & \textbf{Target} & \textbf{TuRBO} & \textbf{ABC} & \textbf{NCS} \\
\midrule
K-S & $0.00$ & \makecell[c]{$0.0368$\\$\pm0.0044$} & \makecell[c]{$0.0483$\\$\pm0.0071$} & \makecell[c]{\bm{$0.0309$}\\\bm{$\pm0.0028$}}  \\
Mean & $7492.59$ & \makecell[c]{$7493.75$\\$\pm1.37$} & \makecell[c]{\bm{$7491.97$}\\\bm{$\pm3.04$}} & \makecell[c]{$7489.52$\\$\pm11.85$}  \\
Std & $57.66$ & \makecell[c]{$59.29$\\$\pm3.96$} & \makecell[c]{$59.26$\\$\pm3.98$} & \makecell[c]{\bm{$58.30$}\\\bm{$\pm3.52$}}  \\
Skewness & $-0.10$ & \makecell[c]{$0.08$\\$\pm0.12$} & \makecell[c]{\bm{$-0.05$}\\\bm{$\pm0.20$}} & \makecell[c]{$0.01$\\$\pm0.14$} \\
Kurtosis & $-0.64$ & \makecell[c]{$0.48$\\$\pm0.22$} & \makecell[c]{$-0.51$\\$\pm0.15$} & \makecell[c]{\bm{$-0.53$}\\\bm{$\pm0.26$}} \\
\bottomrule
\end{tabular}
\end{table}

\subsection{Performance Measures}

\noindent The K-S objective function values obtained by 10 repeated calibrations are compared to verify the performance of NCS, TuRBO, and ABC. First, smaller average of the 10 K-S values indicates stronger capability on estimating multi-modal parameters of the algorithm. Secondly, according to the common stochastic optimization algorithm test standard, the Wilcoxon Rank-Sum Test proves whether the performance differences are statistically significant.

According to the calibration work \cite{21}, the distributional differences between the best simulated data obtained by three algorithms and the target data are compared, including four moments including mean, standard deviation, skewness, and kurtosis of the 1-second level mid-price data (as listed in Table \ref{table2}). For the synthetic data instances, the price trends of the bid/ask order simulated by the three algorithms and the target data are further compared. Since the real data does not have officially publicly accessible order streams, the price trends of the bid/ask orders are not compared.

\begin{figure}[thbp]
\centering
\includegraphics[scale=0.5]{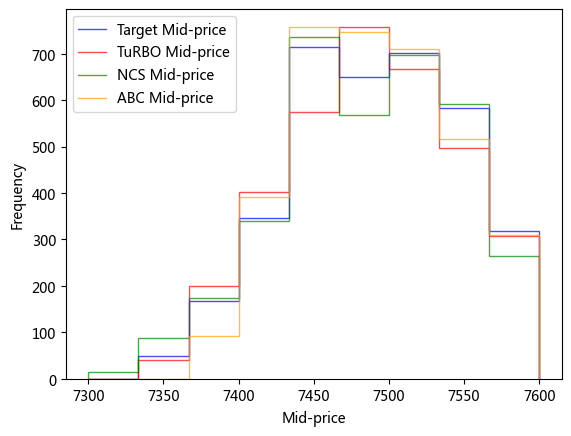}
\caption{The distributions of the mid-price of synthetic data and simulated data generated by ABC, TuRBO and NCS.}
\label{figure7}
\end{figure}

\subsection{Analysis on the Synthetic Data}

\noindent In general, NCS can calibrate the 10 synthetic data instances better than TuRBO and ABC. This can be analyzed in two aspects from Table \ref{table4}. First, the averaged K-S value over 10 repeated runs of NCS is smaller than that of TuRBO and ABC (closer to the target), with an overall improvement of $16.3\%$ and $36.0\%$ respectively. The Wilcoxon rank-sum test also verifies the significance of the advantages of NCS. Secondly, the standard deviations of the 10 K-S values of TuRBO and ABC are 0.0044 and 0.0071. In comparison, the standard deviation of the K-S value of NCS is 0.0028. This means that NCS is more stable, by being beneficial from better diversity through its effective coordination among multiple search processes. This helps NCS has a strong capability on searching for better optima. In fact, in 10 repeated runs, the best and worst K-S values obtained by NCS calibration are 0.0258 and 0.0352, while the best and worst K-S values of TuRBO are 0.0266 and 0.0447 and the corresponding values of ABC are 0.0378 and 0.0528. The above worst results show that NCS indeed has a better search ability in these multi-modal calibration problems.

In particular, the average of K-S value 0.0309 obtained by NCS is smaller than the critical value 0.0320 of the K-S test at a confidence level of 0.95. That means the statistical difference between the synthetic target data and the simulated data generated by NCS is not significant under the K-S test. However, the averaged K-S value 0.0368 and 0.0483 obtained by TuRBO and ABC are both higher than the critical value, which indicates the statistical difference is significant. Hence, the calibration results of NCS are essentially different from those of the compared algorithms. The critical value of K-S test is calculated as $\sqrt{-\frac{(N+n)\ln\frac{\alpha}{2}}{2Nn}}$, where $N$ and $n$ are the length of the calibration data (i.e., 3600) and $\alpha$ is 0.05 at a confidence level of 0.95.

In terms of the indicators of four moments, three algorithms have their own advantages and disadvantages. This further confirms that the traditional calibration objective function, which only compares the moments of target data and simulated data, cannot accurately reflect the differences in data, as well as the merits of the calibration algorithms.

\begin{figure*}[htbp]
\centering
\includegraphics[scale=0.6]{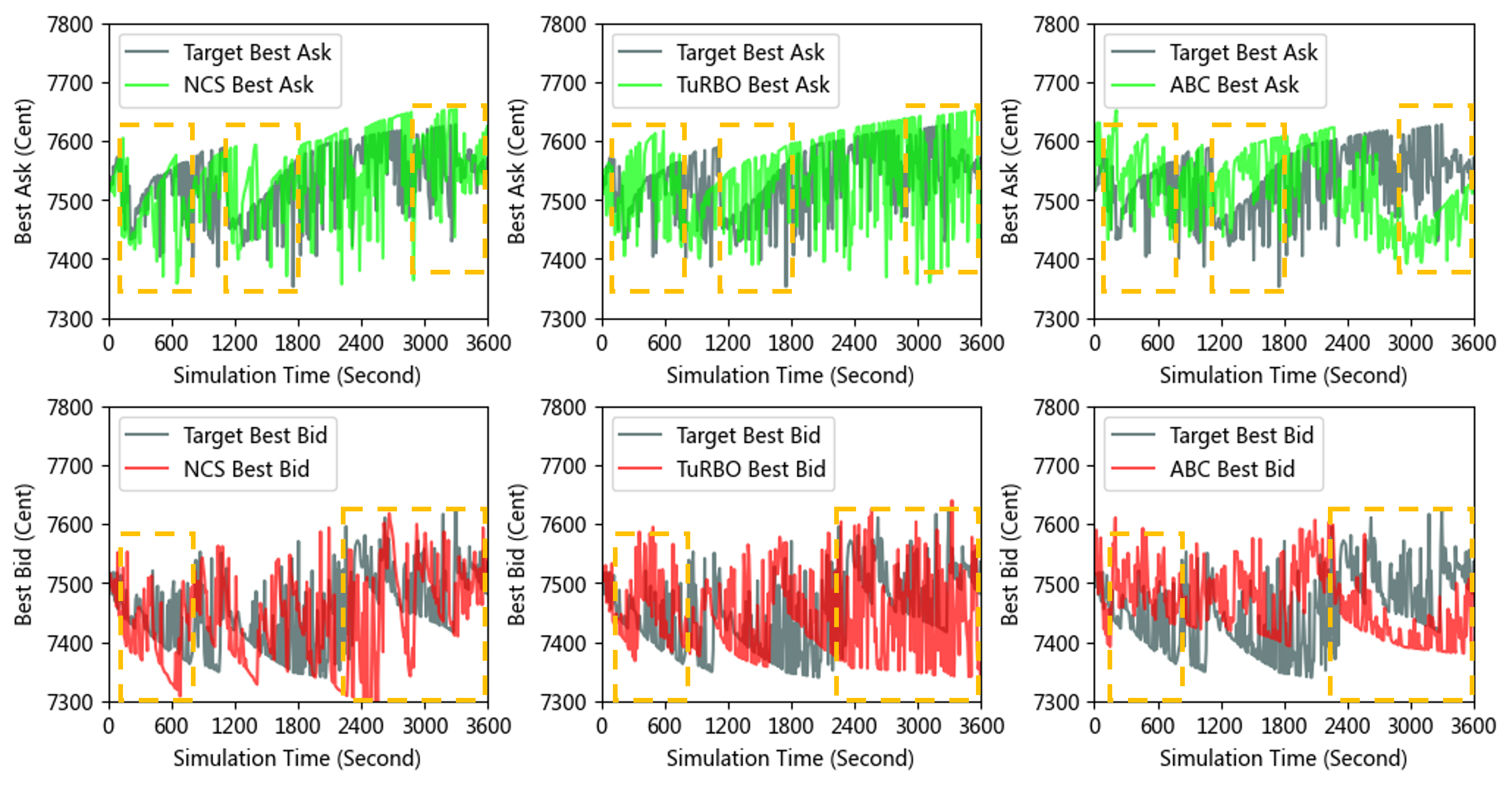}
\caption{The price trends of the best simulated data of NCS (left), TuRBO (middle) and ABC (right). The bid price (in red) and the ask price (in green) of the three algorithms are compared to that of the target Data (in grey).}
\label{figure8}
\end{figure*}

Furthermore, the distributions of the mid-price between the target data and the best simulated data of the three algorithms are compared. For clarity, the frequency histograms of the four mid-price data are depicted. As shown in Fig.\ref{figure7}, the best simulated data of NCS has a smaller difference from the target data, compared with those of TuRBO and ABC. More specifically, the skewness and kurtosis of the mid-price distribution of the best simulated data obtained by NCS are closer to that of the synthetic data.

Finally, the differences between the bid/ask prices in the best simulated data of three algorithms and that in the target data are also compared via visualization. As can be seen from Fig.\ref{figure8}, the simulated data of NCS (upper left and lower left) is significantly superior to the simulated data of the two compared algorithms, since the bid/ask prices of the simulated data of NCS are closer to the target data. The major different areas are highlighted in the yellow dotted box for clarity.

\subsection{Analysis on the Real Market Data}

\noindent As shown in Table \ref{table5}, when NCS is used to calibrate the real 1-second level data, its K-S value is again better than that of TuRBO and ABC. The Wilcoxon rank-sum test is significant as well. On the other hand, the K-S values of NCS, TuRBO, and ABC on real data are all larger than the K-S values calculated on the above synthetic data, which confirms that the real historical data is more difficult to calibrate. One of the main reasons is that the search range of the 6 parameters cannot be determined due to the lake of prior knowledge about the unknown parameter space \cite{35}, where the search space should be set much larger. Compared with the experiments on synthetic data, the selected real data has lower volatility (see the standard deviation indicator value in Table \ref{table5}). All three algorithms can obtain similar standard deviations of the mid-price data by calibrating the model parameters, which shows that the 1-second level real historical data can be calibrated with the studied methods.

\begin{table}[t]
\caption{\textbf{Performance on different indicators of TuRBO, ABC and NCS on real Data with 10 repeated runs}}
\label{table5}
\vspace{8pt}
\centering
\begin{tabular}{ccccc}
\toprule
Index & Target & TuRBO & ABC & NCS \\
\midrule
K-S & $0.00$ & \makecell[c]{$0.0516$\\$\pm0.0044$} & \makecell[c]{$0.0583$\\$\pm0.0088$} & \makecell[c]{\bm{$0.0487$}\\\bm{$\pm0.0060$}}  \\
Mean & $3141.14$ & \makecell[c]{$3141.05$\\$\pm0.11$} & \makecell[c]{\bm{$3141.14$}\\\bm{$\pm0.07$}} & \makecell[c]{$3141.11$\\$\pm0.04$}  \\
Std & $2.21$ & \makecell[c]{\bm{$2.13$}\\\bm{$\pm0.12$}} & \makecell[c]{$2.32$\\$\pm0.22$} & \makecell[c]{$2.05$\\$\pm0.01$}  \\
Skewness & $0.65$ & \makecell[c]{\bm{$0.65$}\\\bm{$\pm0.19$}} & \makecell[c]{$0.79$\\$\pm0.51$} & \makecell[c]{$0.45$\\$\pm0.16$} \\
Kurtosis & $-0.31$ & \makecell[c]{$-0.13$\\$\pm0.37$} & \makecell[c]{$0.03$\\$\pm0.42$} & \makecell[c]{\bm{$-0.27$}\\\bm{$\pm0.15$}} \\
\bottomrule
\end{tabular}
\end{table}

\begin{figure}[t]
\centering
\includegraphics[scale=0.5]{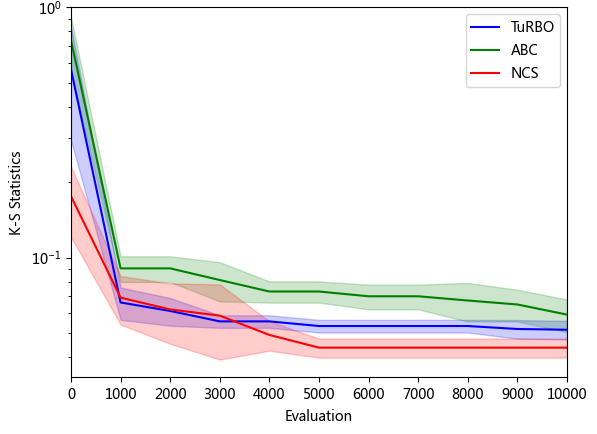}
\caption{The convergence curves (averaged over the 10 independent runs) of NCS, TuRBO and ABC when calibrating the 1-second frequency real data. }
\label{figure9}
\end{figure}

Furthermore, we plot 10 independent calibration curves of NCS, TuRBO, and ABC for real data (Fig.\ref{figure9}). The vertical axis is the best K-S value during the optimization iteration and the horizontal axis is the number of evaluations. Each curve represents the mean and standard deviation of 10 independent runs of each algorithm. In general, the faster the curve drops, the stronger convergence the calibration method performs. From this viewpoint, NCS has a better multi-modal optimization capability than the two compared methods.

\begin{figure}[t]
\centering
\includegraphics[scale=0.5]{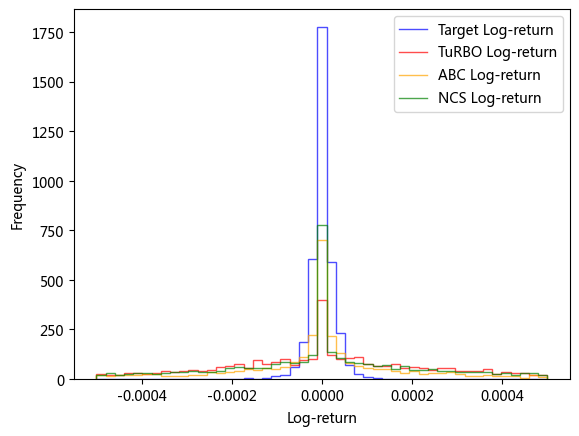}
\caption{Differences in 1-second level mid-price distribution between real data and the simulated data generated by NCS, TuRBO and ABC, respectively.}
\label{figure10}
\end{figure}

Then, we compared the distributional differences of mid-price between the real data and the simulated data corresponding to the best parameters of three algorithms. We statistically analyze the histogram of mid-price data for each of the three data. As shown in Fig.\ref{figure10}, the distribution of the mid-prices of the three data are generally non-dominated that they are closer to the real data within different parts.

\begin{figure}[t]
\centering
\includegraphics[scale=0.5]{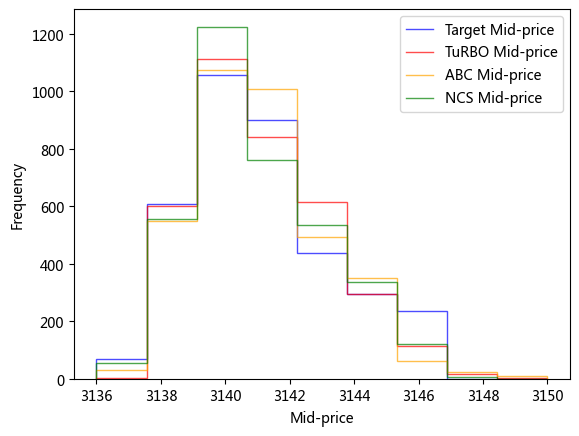}
\caption{Differences in 1-second level log-returns distribution between real data and simulated data generated by NCS, TuRBO and ABC, respectively}
\label{figure11}
\end{figure}

Next, we compare the log-returns of the simulated data and the real data. Log-returns are often used to analyze the financial market data, and is defined as $r_t=logs_t - logs_{t-1}$, which essentially describes the change patterns of the mid- price. Therefore, the log-return sequence $\{r_t\}_{t=2}^T$ can explicitly reflect the data dynamics. As shown in Fig.\ref{figure11}, we plot the difference in the distribution of the 1-second level log-returns of the real data and the simulated data generated by the best parameters of the three algorithms. As seen that, although the highest frequency of the simulated data is quite different from the real data, they all successfully simulated the pattern of the steady-state distribution \cite{10}. Among them, the simulated data of NCS is closer to the log-returns of the real data than the other two methods.

Regarding to the phenomenon that the kurtosis of the two simulated data in Fig.\ref{figure11} is quite different from the real data, this paper believes that the main reason is the absence of the simulation on opening call auction stage. When the real market enters the continuous auction stage, a certain amount of limit order book data has been formed through the opening call auction stage. In this case, new orders with a smaller volume are not likely to cause significant changes in the mid-price. However, the simulation model did not consider the call auction stage. In other words, when entering the continuous auction stage, the limit order book of the simulator contains much less orders so that the middle price is greatly affected by the incoming orders. In this case, the mid-price of the real data generally fluctuates less in the early stage of the simulation, causing its log-return rate closer to 0. Without considering the generation of the initial limit order book by the call auction, the volatility difference between the simulated data and the real data in Fig.\ref{figure11} will exist, which it is difficult for calibration algorithms to solve but needs improvements on the designs of the simulators.

\section{Conclusions and Discussions}

\noindent Given the difficulty of non-identifiability faced by existing FMSs in high-frequency data calibration, this paper starts from the issue of information loss on high-frequency data casued by the existing calibration objective functions. The paper applies visualization to analyze that taking the K-S test as the objective function can significantly reduce non-identifiability, thereby alleviating the issue of information loss. In the process of visualization analysis, this paper further discovers that although the optimal areas of the K-S high-frequency calibration objective function is relatively smaller (higher identifiability), most of these areas are discontinuous. In other words, this turns out to be a multi-modal optimization problem. Traditional black-box optimization algorithms based on a single search process often converge to one optimum, where it is less likely to be the global optimum. This paper proposes an improved NCS algorithm based on adaptive stochastic ranking to solve the above multi-modal optimization problem. Experimental simulations on synthetic data and real data show that the method proposed can significantly improve the fidelity of calibrating high-frequency data. Experiments also find that although the calibration method can indeed benefits better simulating 1-second level real data, the a priori definition of the search range of each parameter appears to be difficult and needs further investigations.

In addition, in the experimental simulations, it is found that the current calibration algorithms require at least 70-80 core hours (i.e., the number of CPU cores multiply the run time on each core) of computational costs to calibrate only one hour of 1-second level data. The main reason is that the process of generating data by the simulator is fairly time-consuming. There are around 5,000 stocks in the Chinese A-share market, which are active for more than 220 days a year and 4 hours a day. If one wants to simulate all the stocks in the market, the computational cost will be unacceptable. Therefore, in the future, apart from further improving the accuracy of calibration, it is necessary to study more computationally efficient calibration methods for FMSs.

\subsection*{Acknowledgment}
This work was supported in part by the National Key Research and Development Program of China under Grant 2022YFA1004102, in part by the National Natural Science Foundation of China (Grant Nos. 62272210, 62250710682, and 62331014), and in part by the Guangdong Major Project of Basic and Applied Basic Research under Grant 2023B0303000010.

\renewcommand\large


\begin{wrapfigure}{l}{25mm} 
    \includegraphics[width=1in,height=1.25in,clip,keepaspectratio]{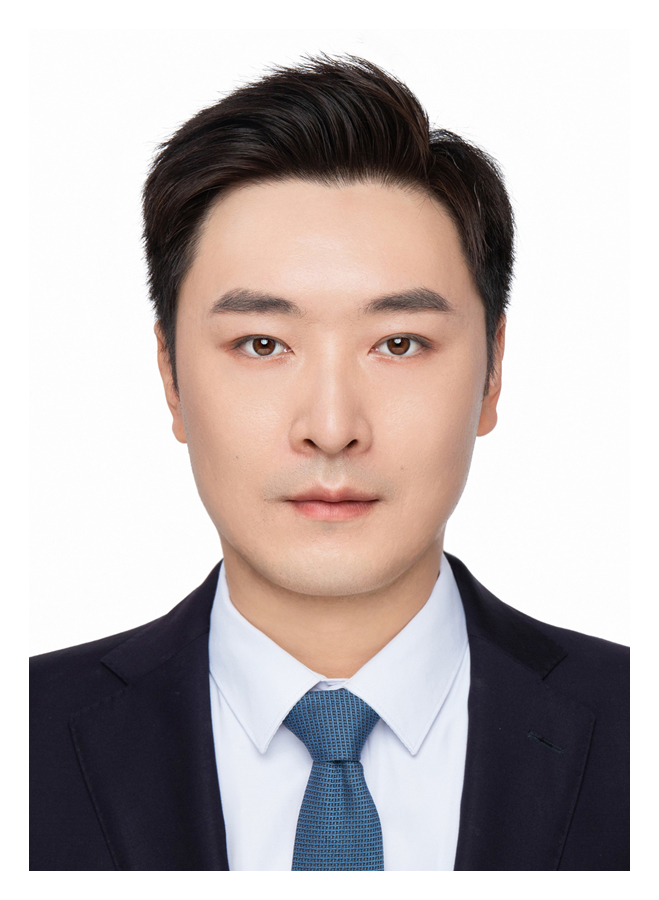}
\end{wrapfigure}\par
\textbf{Peng Yang} received his B.Sc. and Ph.D. degrees in the Department of Computer Science and Technology from University of Science and Technology of China in 2012 and 2017, respectively. From 2017 to 2018, he was a Senior Engineer in Huawei and then he joined Southern University of Science and Technology (SUSTech). He is currently a tenure-track assistant professor jointly in the Department of Statistics and Data Science and the Department of Computer Science and Engineering at SUSTech. His research interests include Evolutionary Computation, Financial Time Series Analysis, and Multi-agent Simulation. He has published over 40 papers in top journals and conferences like TEVC, TCYB, JSAC, TNNLS, TKDE, TASE and NeurIPS. He has also been authorized with 13 invention patents by China, USA, and Germany. He has served as the reviewer for top-tier journals (TEVC, TNNLS, TIE) and the PC member of top conferences (NeurIPS, ICLR, and ICML). He is the vice chair of IEEE CIS Evolutionary Learning Task Force, an executive member of CCF Computational Economics Committee and CCF AI Multi-agent System Committee, and a member of IEEE CIS Evolutionary Computation Technical Committee.\par

\begin{wrapfigure}{l}{25mm} 
    \includegraphics[width=1in,height=1.25in,clip,keepaspectratio]{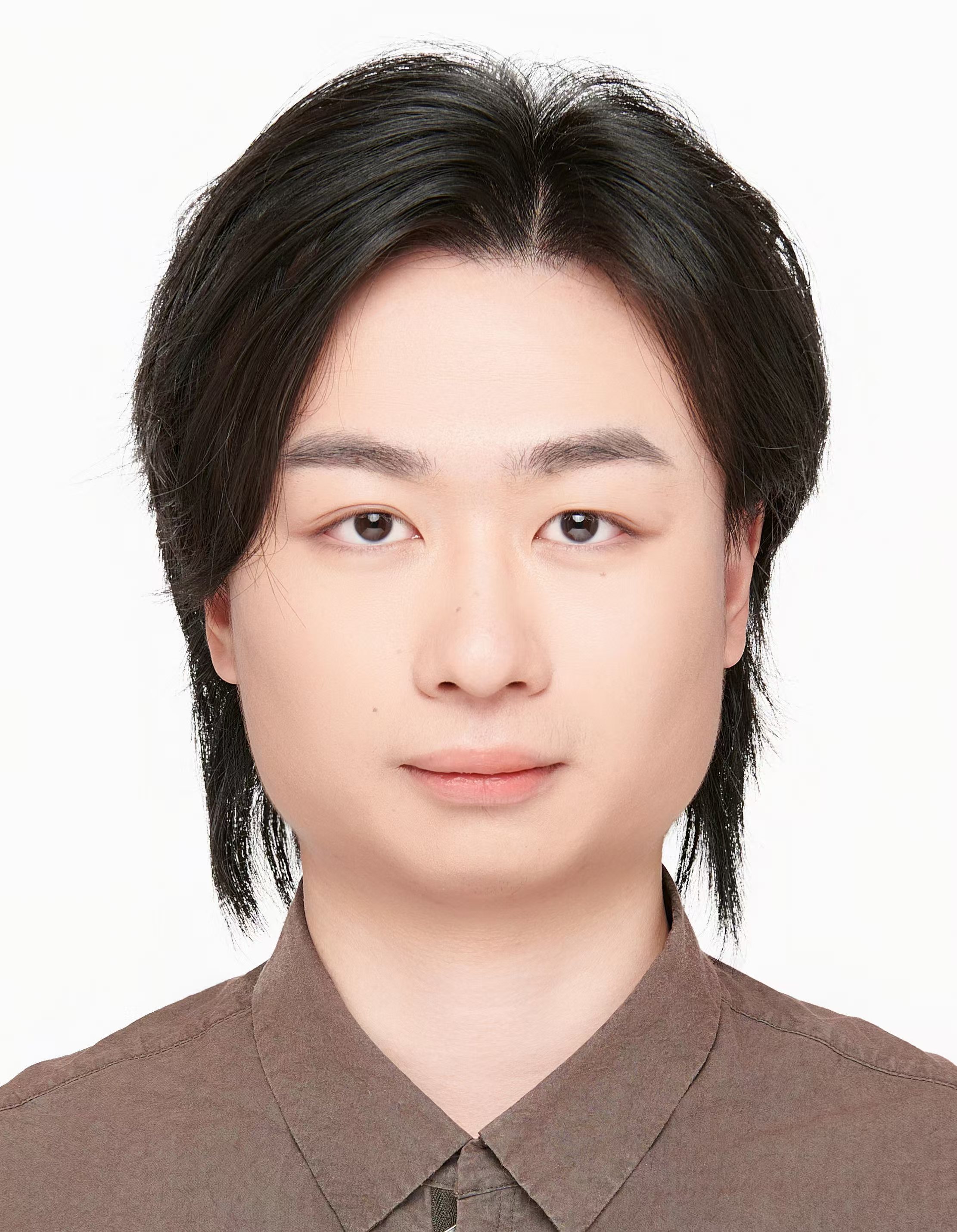}
\end{wrapfigure}\par
\textbf{Junji Ren} received his B.E. degree in the Department of Computer Science and Technology from Southern University of Science and Technology of China in 2024. He is currently pursuing for his M.S. degree of Computer Science at Southern University of Science and Technology. His research interests include evolutionary computation and multi-agent system.\par

\begin{wrapfigure}{l}{25mm} 
    \includegraphics[width=1in,height=1.25in,clip,keepaspectratio]{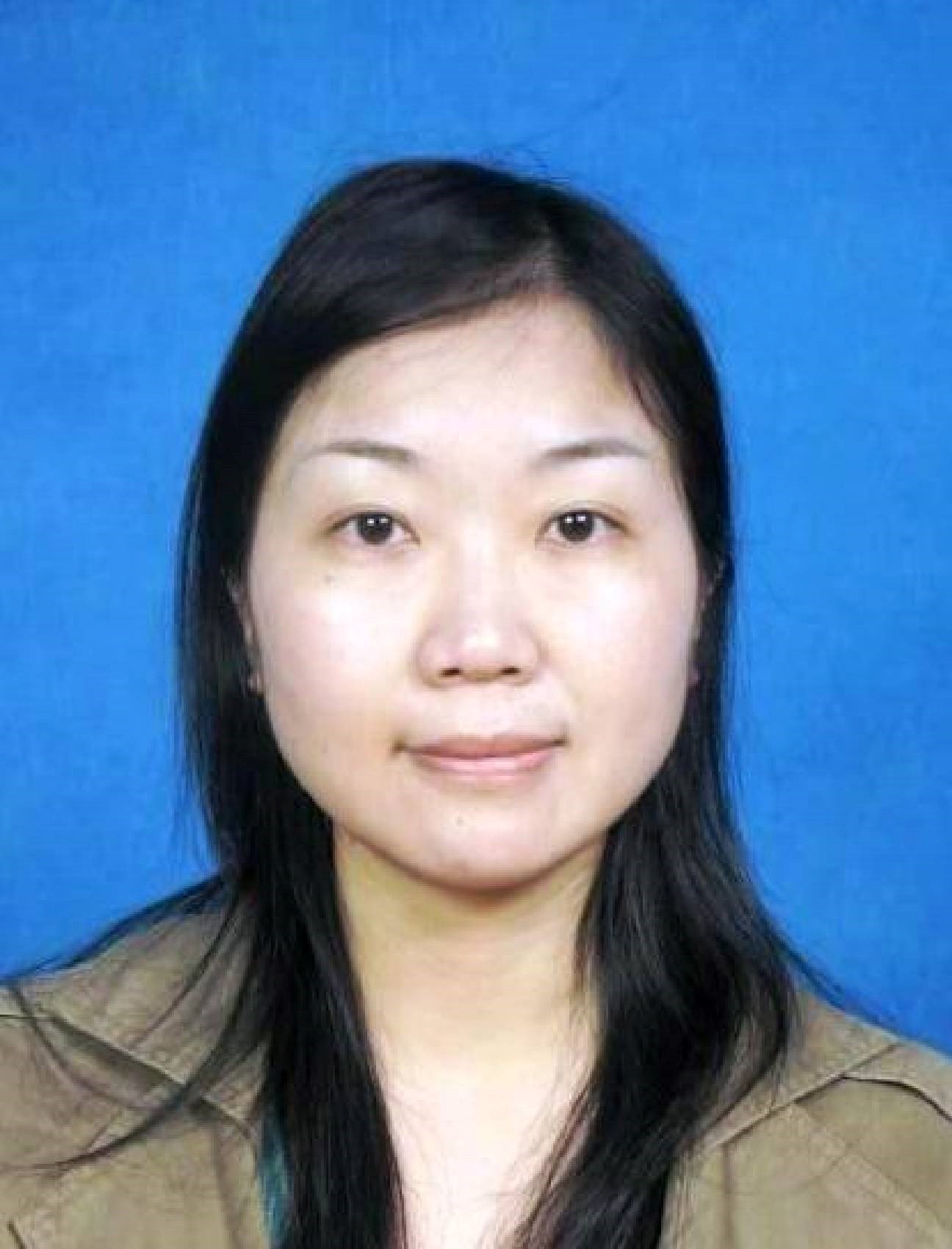}
\end{wrapfigure}\par
\textbf{Feng Wang} received the B.Sc and Ph.D degree in computer science in 2003 and 2008, respectively, both from Wuhan University, Wuhan, China. She is currently a Professor in the School of Computer Science at Wuhan University. Her research interests include evolutionary computation, intelligent information retrieval, and machine learning. She serves as a reviewer for several international journals and conferences.\par

\begin{wrapfigure}{l}{25mm} 
    \includegraphics[width=1in,height=1.25in,clip,keepaspectratio]{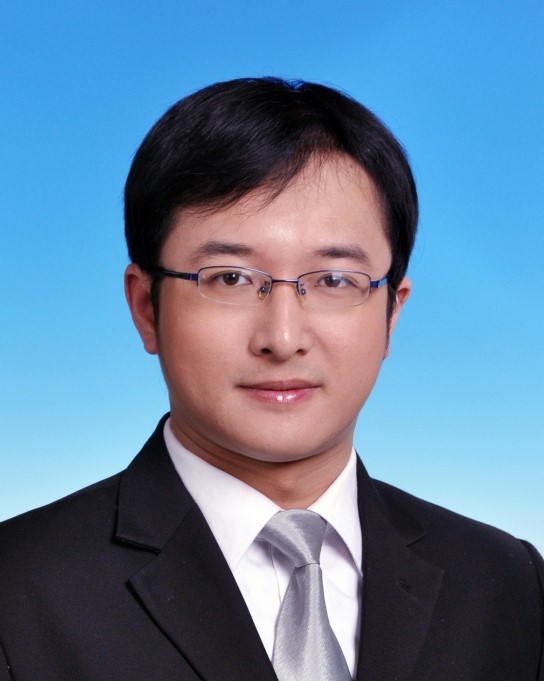}
\end{wrapfigure}\par
\textbf{Ke Tang}
\textbf{(Fellow, IEEE)}
received the Ph.D. degree in computer science from Nanyang Technological University, Singapore, in 2007.

He is a Professor with the Department of Computer Science and Engineering, Southern University of Science and Technology, Shenzhen, China. His research interests mainly include evolutionary computation, machine learning, and their applications, and has authored/coauthored more than 200 papers in these areas.

Dr. Tang was the recipient of the IEEE Computational Intelligence Society Outstanding Early Career Award and the Natural Science Award of Ministry of Education (MOE) of China, and was awarded the Newton Advanced Fellowship (Royal Society) and the Changjiang Professorship (MOE of China). He is an Associate Editor for the IEEE TRANSACTIONS ON EVOLUTIONARY COMPUTATION and a member of editorial boards for a few other journals.\par

\end{document}